\documentclass[conference]{IEEEtran}

\usepackage{cite}
\usepackage{amsmath,amssymb,amsfonts}
\usepackage{algorithmic}
\usepackage{graphicx}
\usepackage{textcomp}
\usepackage{xcolor}
\usepackage{hyperref}
\def\BibTeX{{\rm B\kern-.05em{\sc i\kern-.025em b}\kern-.08em
    T\kern-.1667em\lower.7ex\hbox{E}\kern-.125emX}}

\usepackage{url} 
\usepackage{tcolorbox}
\usepackage{booktabs}
\usepackage{circuitikz}
\usepackage{caption}
\usepackage[table]{xcolor} 
\usepackage{colortbl} 
\definecolor{LightCyan}{rgb}{0.88,1,1}
\definecolor{LightGreen}{rgb}{0.56,0.93,0.56}
\definecolor{LightRed}{rgb}{1,0.6,0.6}

\newcommand{\rmspace}{\vspace{-3ex}}

\begin{document}

\title{Efficient and Reproducible Biomedical Question Answering using Retrieval Augmented Generation}

\author{%
  \IEEEauthorblockN{%
    Linus Stuhlmann\IEEEauthorrefmark{1}, 
    Michael Alexander Saxer\IEEEauthorrefmark{1}, 
    Jonathan Fürst}%
  \IEEEauthorblockA{%
    School of Engineering, Zurich University of Applied Sciences, Winterthur, Switzerland\\
    E-mails: linus.stuhlmann@zhaw.ch, michael.saxer@zhaw.ch jonathan.fuerst@zhaw.ch\\[6pt]
    \IEEEauthorrefmark{1}Equal contribution.
  }%
}

\maketitle

\renewcommand{\thefootnote}{}
\footnotetext{This paper was published at the IEEE Swiss Conference on Data Science (SDS 2025). DOI: \href{https://doi.org/10.1109/SDS66131.2025.00029}{10.1109/SDS66131.2025.00029}.}
\addtocounter{footnote}{-1}

\renewcommand{\thefootnote}{\fnsymbol{footnote}}
\setcounter{footnote}{0}       
\renewcommand{\thefootnote}{\arabic{footnote}} 

\begin{abstract}
Biomedical question-answering (QA) systems require effective retrieval and generation components to ensure accuracy, efficiency, and scalability. This study systematically examines a Retrieval-Augmented Generation (RAG) system for biomedical QA, evaluating retrieval strategies and response time trade-offs. We first assess state-of-the-art retrieval methods, including BM25, BioBERT, MedCPT, and a hybrid approach, alongside common data stores such as Elasticsearch, MongoDB, and FAISS, on a $\approx$ 10\% subset of PubMed (2.4M documents) to measure indexing efficiency, retrieval latency, and retriever performance in the end-to-end RAG system. Based on these insights, we deploy the final RAG system on the full 24M PubMed corpus, comparing different retrievers' impact on overall performance. Evaluations of the retrieval depth show that retrieving 50 documents with BM25 before reranking with MedCPT optimally balances accuracy (0.90), recall (0.90), and response time (1.91s). BM25 retrieval time remains stable (82ms ± 37ms), while MedCPT incurs the main computational cost. These results highlight previously not well-known trade-offs in retrieval depth, efficiency, and scalability for biomedical QA. With open-source code, the system is fully reproducible and extensible.

\end{abstract}


\vspace{-0.3cm}
\section{Introduction}

Large Language Models (LLMs) have demonstrated strong biomedical question-answering (QA) capabilities~\cite{singhal2023expertlevelmedicalquestionanswering}. However, LLMs can produce factual inaccuracies, lack specific domain knowledge, and lack verifiability~\cite{asai2024reliable}. A major concern is \textit{hallucination}, where LLMs generate factually incorrect responses due to their probabilistic nature. These hallucinations, together with a lack of verifiability, are particularly problematic in healthcare, where misinformation can lead to serious consequences. To mitigate these risks, \textit{Retrieval-Augmented Generation (RAG)} systems leverage external knowledge sources at inference time by selecting relevant documents from a data store to enhance accuracy, transparency, and traceability~\cite{tonmoy2024comprehensive}. 

Despite the potential of biomedical RAG systems, existing solutions often suffer from limited scalability, poor reproducibility, and suboptimal retrieval performance on large datasets such as PubMed. 
Existing benchmarks for medical question answering, such as MedExpQA~\cite{alonso2024medexpqa} and MIRAGE~\cite{xiong2024benchmarking}, lack reproducibility and scalable retrieval solutions. 
Most retrieval methods rely on either \textit{sparse} bag-of-words vectors such as BM25~\cite{robertson2009probabilistic} or \textit{dense} vectors created through transformer-based models such as BioBERT~\cite{Lee_2019} and MedCPT~\cite{Jin2023MedCPT}.
However, \textbf{hybrid approaches that integrate both techniques remain under-investigated, especially from a system perspective}: the inherent trade-offs between retrieval strategies, their indexing and response times, and the resulting generator accuracy are crucial for practical RAG applications and have been largely unexplored. Hybrid retrieval methods combine the strengths of sparse and dense retrieval: a \textit{probabilistic retriever} (e.g., BM25) efficiently reduces the search space by filtering a large corpus, while a \textit{neural reranker} (e.g., MedCPT’s cross-encoder) refines document rankings based on semantic relevance.  This two-step approach balances computational efficiency and retrieval precision, mitigating the limitations of stand-alone methods. Although hybrid retrieval has been explored in general NLP tasks~\cite{ma2020hybrid}, \textit{its application in large-scale biomedical QA remains limited, particularly in real-world implementations}. Developing an effective biomedical QA system requires addressing several challenges:
(i) \textbf{Efficient Retrieval at Scale}: Processing millions of biomedical documents under reasonable \textit{indexing times} while maintaining \textit{low-latency retrieval}; (ii) \textbf{Relevance Optimization}: Improving \textit{document ranking} by integrating lexical retrieval with neural reranking; (iii) \textbf{Context Integration}: Structuring retrieved documents effectively to generate \textit{factually accurate and verifiable responses}. This work presents a \textbf{scalable and reproducible} RAG system for biomedical QA, systematically evaluating hybrid retrieval strategies. The key contributions include:

\begin{itemize}
    \item \textbf{Hybrid Retrieval Approach}: A two-stage retrieval pipeline that integrates BM25 (lexical retrieval) with MedCPT’s cross-encoder (semantic reranking), improving recall and precision.
    \item \textbf{Scalability and Performance Analysis}: Comparative evaluation of three common methods and systems,\\ MongoDB, Elasticsearch and FAISS, for large-scale document retrieval efficiency.
    \item \textbf{Reproducibility and Transparency}: Explicit citation of retrieved documents using PubMed IDs to ensure traceability in biomedical QA.
\end{itemize}
\newpage
This work advances scalable and reproducible biomedical QA systems, enhancing their real-world applicability in clinical and research environments. All our code is open-sourced.\footnote{\url{https://github.com/slinusc/medical_RAG_system}}


\section{Biomedical Question-Answering with RAG}
\label{sec:BQA-RAG}
\begin{figure}[ht]
\centering
\includegraphics[width=0.8\columnwidth]{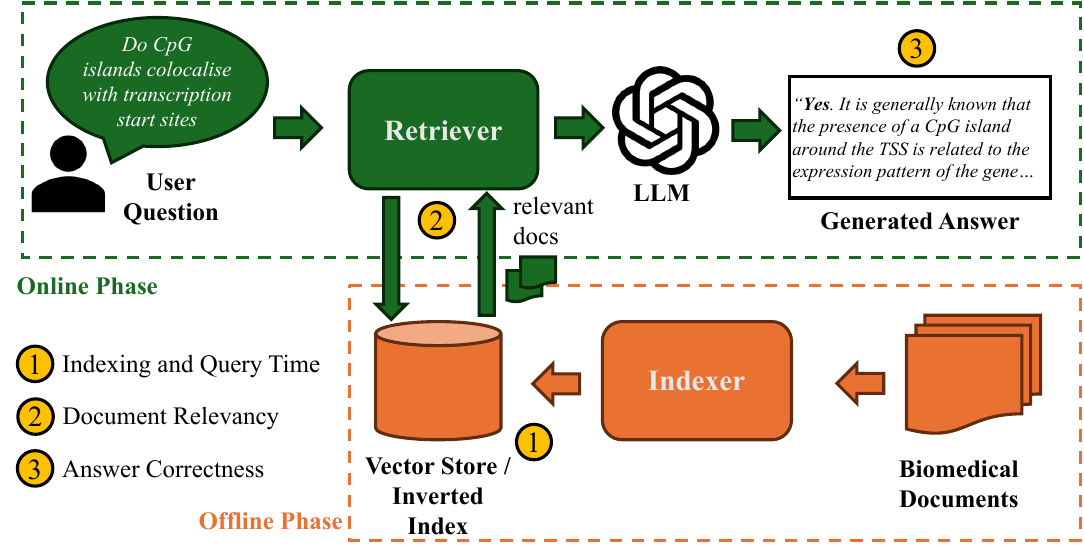}
\caption{Biomedical Question Answering with Retrieval-Augmented Generation (RAG). Offline phase: biomedical documents are processed, indexed, and stored in a vector store. Online phase: users ask questions, for which a retriever retrieves relevant documents that are appended with the question and fed to an LLM. Based on the question and context, the LLM generates an answer the PubMed IDs it used.}
\label{fig:RAG_workflow}
\end{figure}
Retrieval-Augmented Generation enhances the capabilities of LLMs by incorporating external data and grounding \\ responses in verifiable and up-to-date information. This ensures that outputs incorporate relevant biomedical knowledge from sources such as PubMed~\cite{lewis2020retrieval}, improving accuracy and transparency.
Figure~\ref{fig:RAG_workflow} illustrates our biomedical QA system based on RAG, which consists of two main phases: an offline phase for indexing biomedical literature and an online phase for retrieving relevant documents and generating responses.

In the offline phase, biomedical documents are preprocessed and indexed into a vector store (dense vectors) and/or an inverted index (sparse vectors) for efficient retrieval. The choices directly affect retrieval speed and system scalability.

During the online phase, users submit biomedical queries, which the retriever processes to fetch relevant documents. These retrieved documents, along with the query, are provided as context to an LLM, ensuring responses remain grounded in authoritative biomedical sources. The used sources are cited (PMIDs) and provided as references to users. In this setting, system performance should be evaluated based on three key aspects, as highlighted in Figure~\ref{fig:RAG_workflow}: (1) indexing and query time, which measures retrieval efficiency; (2) relevance of retrieved documents, ensuring the most informative sources are selected; and (3) answer correctness, verifying that the LLM-generated response aligns with biomedical evidence. Evaluating and optimizing these factors improves the reliability and transparency of biomedical QA systems.

\section{Experimental Evaluation}

We evaluate the efficiency and effectiveness of different retrieval and text-generation methods for biomedical question-answering (QA). Our experiments focus on selecting optimal components for document retrieval, text generation, and overall system performance following the three key aspects from Section~\ref{sec:BQA-RAG} and using a common biomedical QA benchmark.

\subsection{Experimental Setup}

\textbf{Datasets.} We evaluate our biomedical RAG system on the BIOASQ~\cite{krithara2023bioasq} QA benchmark, which builds on the PubMed database~\cite{xiong2024benchmarking}.
Specifically, we use a 10\% randomly sampled subset of 2.4M biomedical papers for the component analysis, while for the final system, we use the entire dataset of \textbf{24M}. Each entry includes a PubMed ID (PMID), title, and abstract, with an average abstract length of 296 tokens.
We evaluate our system using the Task-B dataset, which contains expert-annotated questions paired with supporting PubMed IDs (PMIDs).
To ensure answerability within our PubMed subset, we first exclude factoid and list questions, which often require full-text access, making evaluation less precise.
Second, we retain only questions with at least one PMID in our dataset to ensure they can be answered using our subset.



\textbf{Indexing and Retrieval Systems.}
We compare three storage and query systems: Elasticsearch, FAISS, and MongoDB.

MongoDB\footnote{\url{https://www.mongodb.com/}}, a NoSQL document database, supports full-text search with TF-IDF-like scoring in its self-hosted version. While BM25 ranking is available in MongoDB Atlas Search, it is a cloud-only service and was not used.

Elasticsearch\footnote{\url{https://www.elastic.co/elasticsearch}}, built on Apache Lucene, uses BM25 ranking and inverted indexing for efficient text-based retrieval.

FAISS (Facebook AI Similarity Search) optimizes dense vector similarity search, commonly used in NLP and recommendation systems \cite{douze2024FAISS}. We deployed FAISS using a Flask-based server with a FlatL2 index for exhaustive search.

\colorbox{blue!10}{Metrics:} We evaluate \textit{indexing speed} and \textit{response time} on 2.4M PubMed papers to determine the best trade-off between efficiency and retrieval performance.

\textbf{Retrieval Methods.}
Based on recall and precision, we evaluate four retrieval methods—BM25, BioBERT, MedCPT, and a hybrid approach (BM25 + MedCPT).

BM25~\cite{robertson1994okapi} is a ranking algorithm that improves upon TF-IDF~\cite{sparck1972} by incorporating term frequency, document length normalization, and inverse document frequency. It ranks documents based on query relevance using a probabilistic scoring function. We implemented BM25 in Elasticsearch, with stopword removal for improved efficiency.
BioBERT \cite{Lee_2019} is a domain-specific adaptation of BERT \cite{devlin2019bert}, pre-trained on PubMed abstracts and PMC articles to enhance biomedical text understanding. We use BioBERT to encode PubMed abstracts into semantic vectors via FAISS, computing document-query similarity with squared Euclidean distance.
%
MedCPT \cite{Jin2023MedCPT} is a contrastive learning-based retriever trained on 255M PubMed query-article interactions. It consists of a query encoder, document encoder, and a cross-encoder reranker. The cross-encoder refines retrieval results by reranking top candidates based on query-document contextual interactions. We use MedCPT to encode 2.4M abstracts. We filter results based on positive relevance scores.
The Hybrid Retriever integrates BM25 and MedCPT for enhanced retrieval performance. BM25 first ranks a broad set of documents in Elasticsearch, after which MedCPT's cross-encoder reranks the top-\(k\) results. This combination leverages BM25’s efficiency and MedCPT’s semantic understanding to improve recall and precision.

\colorbox{blue!10}{Metrics:} \textit{We assess how well each method retrieves relevant documents}. Since \textit{recall} is critical for ensuring comprehensive retrieval, we prioritize methods that maximize relevant document retrieval while maintaining high \textit{precision}.

\textbf{Text Generation.}
For text generation, we experiment with different prompting strategies for OpenAI’s GPT-3.5-turbo (API version May 2024, temperature=0), ensuring that generated responses are accurate and contextually relevant.
Given the biomedical domain's strict accuracy requirements, we focus on structured prompts that enhance factual consistency.
We experimented with multiple prompting approaches, following best practices in medical NLP~\cite{mesko2023prompt, xiong2024benchmarking}. Due to resource constraints, we evaluated GPT-3.5, with limited testing of GPT-4. Observations showed no significant differences in output quality.
%
As illustrated in Figure \ref{fig:promt}, our final prompt consists of three components: (1) a system prompt with task-specific instructions, (2) a user query, and (3) retrieved documents with PubMed IDs (PMIDs), titles and content.

\colorbox{blue!10}{Metrics:} For text generation, we evaluate \textit{answer correctness} in terms of \textit{accuracy}, \textit{recall}, \textit{precision}, and \textit{F1 score}.

\begin{figure}[]
\centering

\begin{tcolorbox}[colframe=black, colback=white, sharp corners=south, title=Prompt Template for Medical QA,
boxsep=0.5mm,
    left=1mm,
    right=1mm,
    top=1mm,
    bottom=1mm]
\footnotesize
System Prompt: You are a scientific medical assistant designed to synthesize responses from specific medical documents. Only use the information provided in the documents to answer questions. The first documents should be the most relevant. Do not use any other information except for the documents provided. When answering questions, always format your response as a JSON object with fields for 'response', 'used\_PMIDs'. Cite all PMIDs your response is based on in the 'used\_PMIDs' field. Please think step-by-step before answering questions and provide the most accurate response possible. Provide your answer to the question in the 'response' field.

\vspace{0.5ex}
User Prompt:
Answer the following question: {...}

\vspace{0.5ex}
Context Prompt:
Here are the documents:

\vspace{-0.5ex}
\begin{verbatim}
"doc1": {
    "PMID": {...},
    "title": {...},
    "content": {...}
    "relevance_score": {...}
}, ...
\end{verbatim}

\end{tcolorbox}
\caption{Prompting approach for biomedical QA.}
\label{fig:promt}
\rmspace
\end{figure}

\subsection{Indexing and Query Time}

Table \ref{table:search_performance} summarizes the performance of Elasticsearch, FAISS, and MongoDB. Elasticsearch excels in full-text retrieval but is less efficient for semantic vector search, which FAISS optimizes for. However, due to their complex data management and indexing mechanisms, MongoDB and Elasticsearch exhibit the slowest indexing speeds.

MongoDB, while providing a flexible NoSQL document storage solution, uses TF-IDF-based text ranking in its self-hosted version, which leads to significantly slower query response times compared to Elasticsearch and FAISS. The self-hosted MongoDB lacks efficient semantic retrieval, limiting its effectiveness in large-scale biomedical QA.

Based on these results, we selected Elasticsearch for full-text retrieval and FAISS for semantic vector search. Despite its slower indexing speed, Elasticsearch provides a robust text-based search framework, while FAISS offers superior response times for vector-based queries.

\begin{table}[hb] 
    \caption{Performance metrics for different search methods.}
    \label{table:search_performance}
    \centering
    \scalebox{.77}{
    \begin{tabular}{|l|l|l|l|l|}
        \hline
        \rowcolor{gray!20}
        \textbf{Method} & \textbf{Type} & \textbf{Index} & \textbf{Response Time} & \textbf{Indexing Speed} \\
        \hline
        MongoDB & Sparse & TF-IDF & \cellcolor{red!40}26.4 s ± 1.72 s & \cellcolor{green!40}10.41 min \\
        \hline
        Elasticsearch & Sparse & BM25 & \cellcolor{green!40}82 ms ± 37 ms & \cellcolor{red!40}156 min \\
        \hline
        Elasticsearch & Dense & KNN & \cellcolor{red!40}24.6 s ± 1.23 s & \cellcolor{red!40}171 min \\
        \hline
        FAISS & Dense & L2 Distance & \cellcolor{green!10}657 ms ± 127 ms & \cellcolor{green!10}41 min \\
        \hline
    \end{tabular}
    }
    \rmspace
\end{table}

\subsection{Document Relevancy}

Table \ref{table:recall_retreiver} summarizes the retrievers' performance. The Hybrid Retriever achieved the highest recall (0.567), balancing efficiency and accuracy. BM25 exhibited strong precision but lower recall. MedCPT improved semantic retrieval but underperformed in recall, while BioBERT had the weakest results due to a lack of fine-tuning for question-answering tasks.
Note that a low recall score does not necessarily indicate incorrect retrieval; rather, it means that the retrieved documents may not be included in the BioASQ-curated set.

\begin{table}[hb]
    \caption{Performance comparison of different retrievers.}
    \label{table:recall_retreiver}
    \centering
    \scalebox{0.77}{
\begin{tabular}{|l|c|c|c|}
        \hline
        \rowcolor{gray!20}
        \textbf{Retriever} & \textbf{Vector Type} & \textbf{Recall} & \textbf{Precision}\\
        \hline
        Hybrid Retriever & Hybrid & \cellcolor{green!40}0.567  & \cellcolor{green!10}0.319  \\
        \hline
        BM25 & Sparse & \cellcolor{green!40}0.537 & \cellcolor{green!10}0.322 \\
        \hline
        MedCPT & Dense & \cellcolor{red!10}0.273 & \cellcolor{red!10}0.205\\
        \hline
        BioBERT & Dense & \cellcolor{red!40}0.07 & \cellcolor{red!40}0.07\\
        \hline
    \end{tabular}
    }
    \rmspace
\end{table}

\subsection{Answer Correctness of the RAG System (End-to-End)}

%
For \textbf{BM25}, the query is processed using term-based retrieval, ranking documents based on query term occurrence. The top \(k\) ranked documents are embedded into the LLM context for response generation.
For \textbf{MedCPT}, the query is encoded into a vector and compared against document embeddings for similarity search. The retrieved documents are reranked by a cross-encoder, and only those with positive relevance scores are used for response generation.
For \textbf{Hybrid Retrieval}, BM25 first retrieves \(k\) candidate documents, which are then reranked by MedCPT’s cross-encoder. Only relevant documents are passed to the LLM. Our results show that the \textit{hybrid retriever} achieves the \textit{best answer correctness} on all metrics (Table \ref{fig:performance_models}).


\begin{table}[hb]
    \caption{Performance metrics of the end-to-end RAG system using different retrievers.}
    \label{fig:performance_models}
    \centering
    \scalebox{0.77}{
    \begin{tabular}{|l|c|c|c|c|}
        \hline
        \rowcolor{gray!20}
        \textbf{RAG with Retriever} & \textbf{Accuracy} & \textbf{Recall} & \textbf{Precision} & \textbf{F1 Score} \\
        \hline
        GPT-3.5 / Hybrid Retriever & \cellcolor{green!40}0.86 & \cellcolor{green!40}0.86 & \cellcolor{green!40}0.89 & \cellcolor{green!40}0.86 \\
        \hline
        GPT-3.5 / MedCPT & \cellcolor{green!10}0.83 & \cellcolor{green!10}0.83 & \cellcolor{green!40}0.86 & \cellcolor{green!10}0.84 \\
        \hline
        GPT-3.5 / BM25 & 
        \cellcolor{red!10}0.72 & \cellcolor{red!10}0.72 & \cellcolor{green!10}0.83 & \cellcolor{red!10}0.74 \\
        \hline
        GPT-3.5 / BioBERT & \cellcolor{red!40}0.63 & \cellcolor{red!40}0.63 & \cellcolor{green!10}0.85 & \cellcolor{red!40}0.67 \\
        \hline
    \end{tabular}
    }
    \rmspace
\end{table}

\section{Evaluation of the Final System}

After selecting the most efficient and effective components, we evaluate the final hybrid RAG system on the full 24M-document PubMed corpus for \textit{retrieval effectiveness}, \textit{response time}, and \textit{answer correctness}.

\subsection{Effect of Retrieval Depth on Performance}

To evaluate the impact of retrieval depth on performance, we experimented with different configurations of BM25 retrieval, varying the number of initially retrieved documents while keeping the reranking step fixed at the top 10 (Table \ref{table:retrieval_performance}).

\begin{table}[ht]
    \caption{Comparison for different retrieval depths  (BM25), with reranking applied to the top 10 documents.}
    \label{table:retrieval_performance}
    \centering
    \scalebox{0.75}{
    \begin{tabular}{|c|c|c|c|c|c|c|}
        \hline
        \rowcolor{gray!20}
        Docs & Accuracy & Recall & Precision & F1 Score & Retrieval Time (s) & Total Time (s) \\

        \hline
        20  & \cellcolor{yellow!10}
0.89 &\cellcolor{yellow!10}
 0.88 &\cellcolor{green!40}
 0.89 & \cellcolor{yellow!10}
0.88 &\cellcolor{green!40}
 0.39 $\pm$ 0.07 &\cellcolor{green!40}
 1.52 $\pm$ 0.42 \\
        \hline
        50  &\cellcolor{green!40}
0.90 &\cellcolor{green!40}
 0.90 &\cellcolor{green!40}
 0.89 &\cellcolor{green!40}
 0.90 & \cellcolor{yellow!10}
0.82 $\pm$ 0.13 & \cellcolor{yellow!10}
1.91 $\pm$ 0.36 \\
        \hline
        100 &\cellcolor{red!40}
 0.87 &\cellcolor{red!40}
 0.87 & \cellcolor{red!40}
0.88 &\cellcolor{red!40}
 0.87 &\cellcolor{red!40}
 1.54 $\pm$ 0.16 &\cellcolor{red!40}
 2.62 $\pm$ 0.44 \\
        \hline
    \end{tabular}
    }
    \rmspace
\end{table}
\cellcolor{green!40}
\cellcolor{red!40}
\cellcolor{yellow!10}



\subsection{Analysis of Retrieval Depth Trade-offs}

Elasticsearch BM25 retrieval has an average response time of $82\pm 37$ms, which remains constant across all retrieval depths since it ranks all documents regardless of how many are later passed to reranking. The primary factor affecting response time is the cross-encoder reranking step using MedCPT, which processes a subset of the retrieved documents and incurs additional computational overhead.

Increasing the number of retrieved documents leads to marginal accuracy improvements but significantly increases the rerank time. Retrieving 50 documents before reranking yields the best accuracy (0.90) and F1 score (0.90) while keeping response time manageable at 1.91 seconds. However, retrieving 100 documents leads to a drop in accuracy (0.87) and an increase in total response time to 2.62 seconds, suggesting diminishing returns beyond 50 documents.

The text generation phase relies on the OpenAI API, which introduces additional latency. The mean response time for generation is 1.07 seconds, with a standard deviation of 0.41 seconds. Since the generation time remains stable across configurations, the overall system latency is primarily determined by the retrieval depth and reranking time.

These results demonstrate that increasing the number of retrieved documents beyond a certain threshold does not necessarily improve system performance. Instead, balancing retrieval depth with reranking efficiency is critical for real-world biomedical question-answering applications.

\section{Conclusion and Future Directions}

Biomedical question-answering (QA) systems require both efficient retrieval and generation components for accuracy and scalability. This study examines a Retrieval-Augmented Generation (RAG) system for biomedical QA, evaluating retrieval strategies and response time trade-offs.

We assess retrieval methods, including BM25, BioBERT, MedCPT, and a hybrid approach, alongside data stores such as Elasticsearch, MongoDB, and FAISS.
Despite strong performance, some limitations remain. The reliance on OpenAI’s GPT-3.5 for text generation poses reproducibility challenges due to model updates and API latency. Additionally, retriever and database system evaluations remain limited, requiring broader comparisons.

Future work should explore additional retrievers
and evaluate alternative databases
for indexing efficiency. Efforts should also focus on retrieval optimization, integrating open-source LLMs, and enabling real-time biomedical applications. Our work highlights trade-offs in retrieval depth, efficiency, and scalability. The system is fully reproducible and extensible, supporting future advancements in retrieval and model integration for research and clinical applications.

\bibliographystyle{IEEEtran} 
\bibliography{bibliography}

\end{document}